\documentclass[debug]{rmaa}

	\usepackage{paralist}

	\usepackage{psfrag,color}

	\usepackage[latin1]{inputenc}

	\usepackage{pdflscape}
	\usepackage{rotating}
	\usepackage{graphicx}
	\usepackage{epstopdf}
	\usepackage{epsfig}
	\usepackage{booktabs}
	\usepackage{dcolumn}

	\title{the IRAS 08589$-$4714 star-forming region} 

	\author{H. P. Salda\~no\altaffilmark{1,2},
	J. V\'asquez\altaffilmark{2,3,4}, 
	M. G\'omez\altaffilmark{1,2},
	C. E. Cappa\altaffilmark{2,3,4},
	N. Duronea\altaffilmark{2,3} and
	M. Rubio\altaffilmark{5}
	}

	\altaffiltext{1}{Observatorio Astron\'omico, Universidad Nacional de C\'ordoba,
	C\'ordoba, Argentina.}

	\altaffiltext{2}{CONICET, Consejo Nacional de Investigaciones Cient\'\i ficas y T\'ecnicas,
	Argentina.}

	\altaffiltext{3}{Instituto Argentino de Radioastronom\'ia, CONICET, CCT La Plata,
	Villa Elisa, Argentina.}

	\altaffiltext{4}{ Facultad de Ciencias Astron\'omicas y Geof\'isicas,
	Universidad Nacional de la Plata, La Plata, Argentina.}

	\altaffiltext{5}{Departamento de Astronom\'\i a, Universidad de Chile,
	Santiago de Chile, Chile.}

	\shortauthor{Salda\~no et al.}
	\shorttitle{IRAS 08589$-$4714 star-forming region}

	\fulladdresses{

	\item Cappa, C. E. and V\'asquez, J.: CONICET, Consejo Nacional de Investigaciones Cient\'\i ficas
	y T\'ecnicas, Argentina, Instituto Argentino de Radioastronom\'ia, CONICET, CCT La Plata, C.C.5, 1894,
	Villa Elisa, Argentina, Facultad de Ciencias Astron\'omicas y Geof\'isicas, Universidad Nacional de la
	Plata, Paseo del Bosque s/n, 1900, La Plata, Argentina (ccappa@fcaglp.unlp.edu.ar, pete@iar.unlp.edu.ar).
	\item Duronea N.: Consejo Nacional de Investigaciones Cient\'\i ficas y T\'ecnicas, Argentina, Instituto
	Argentino de Radioastronom\'ia, CONICET, CCT La Plata, C.C.5, 1894, Villa Elisa, Argentina
	(duronea@iar.unlp.edu.ar).
	\item Rubio, M.: Departamento de Astronom\'\i a, Universidad de Chile, Casilla 36, Santiago de Chile,
	Chile (mrubio@das.uchile.cl).
	\item Salda\~no, H. P. and G\'omez, M.: Observatorio Astron\'omico, Universidad Nacional de C\'ordoba, Laprida 854, 
	5000 C\'ordoba, Argentina (hugosaldanio, mercedes@oac.uncor.edu).}

	\listofauthors{H. P. Salda\~no, J. V\'asquez, C. E. Cappa, M. G\'omez, N. Duronea, \& M. Rubio}
	\indexauthor{Salda\~no, H. P.}
	\indexauthor{Vasquez, J.}
	\indexauthor{Cappa, C. E.}
	\indexauthor{G\'omez, M.}
	\indexauthor{Duronea, N.}
	\indexauthor{Rubio, M.}

	\abstract{We present an analysis of the IRAS\,08589$-$4714 star-forming region.
	This region harbors candidate young stellar objects identified in the WISE and 
        {\textit{Herschel}} images using { color index
	criteria} and spectral energy distributions (SEDs). The SEDs of some of the infrared sources and
	the 70 $\mu$m radial intensity profile of the brightest source (IRS 1) { are modeled 
	using the one-dimensional radiative transfer DUSTY code}.  For these objects, we estimate the envelope
	masses, sizes, densities, and luminosities which suggest that they are very young, massive and
	luminous objects at early stages of the formation process. 
        Color-color diagrams in the bands of WISE and 2MASS are used to identify potential 
        young objects in the region. Those identified in the bands of WISE would be contaminated
        by the emission of PAHs.
	We use the emission distribution in the infrared at 70 and 160  $\mu$m, to estimate the dust
        temperature gradient. This suggests that the nearby massive star-forming region RCW\,\,38,
        located $\sim$ 10 pc of the
        IRAS source position may be contributing to the photodissociation of the molecular
	gas and to the heating of the interstellar dust in the environs of the IRAS source.}

	\resumen{Se presenta un análisis de la región de formación de estrellas masivas IRAS 08589$-$4714.
        Esta región alberga candidatos a objetos estelares jóvenes, {los cuales fueron} identificados en las imágenes de WISE y 
        {\textit{Herschel}} empleando criterios { basados en índices de color} y distribuciones 
        espectrales de energía (SEDs). Las SEDs de algunas de las fuentes y el perfil radial de intensidad 
        de la fuente más brillante en la banda de 70 $\mu$m de {\textit{Herschel}} (IRS 1) son modelados mediante el 
        { código unidimensional de transporte radiativo DUSTY}. Para estos objetos, se estiman las masas de las
        envolventes, los tamaños y densidades de las mismas y las luminosidades. Estos parámetros indican que
        se trata de objetos muy jóvenes, masivos y luminosos en las primeras etapas del proceso de formación.
        Se emplean los diagramas color-color en las bandas de WISE y de 2MASS para indentificar potenciales
        objetos jóvenes en la región. Aquellos identificados en las bandas de WISE estarían contaminados 
        por la emisión de PAHs.
        Se emplea la distribución de la emisión en el infrarrojo en 70 y 160 $\mu$m, para estimar el gradiente
	de temperatura del polvo. Esto indica que la región de formación estelar masiva cercana RCW\,\,38,
        localizada $\sim$ 10\,\,pc de la posici\'on de la fuente IRAS, podría
	contribuir a la fotodisociación del gas molecular
	y al calentamiento del polvo interestelar en las inmediaciones de la fuente IRAS. }

	\addkeyword{Stars: circumstellar matter}
	\addkeyword{Stars: formation}
	\addkeyword{Stars: massive}
	\addkeyword{ISM: dust, extinction}
	\addkeyword{ISM: individual objects (IRAS 08589$-$4714)}
	\addkeyword{Infrared: stars}

\begin{document}
	\maketitle

	\section{Introduction}
	\label{sec:intro}

	Massive stars are believed to form in dense  ($n$ $\sim 10^{3-8}$ cm$^{-3}$),
	cold ($T$\,\,$<$\,\,30\,\,K), very opaque ($\tau_{\rm 100}\footnote{Optical
	depth at 100 $\mu$m.} \sim$ 1\,--\,6) and massive ($M$ $\sim$ 8\,--\,2000 $M_{\sun}$)
	regions,  known as pre-stellar cores
	\citep{2002A&A...390.1001S,2009A&A...494..157D,2010A&A...516A.102C,2014ApJ...787..113B}.
	\citet{2009ApJS..181..360C} analyzed a significant number of objects having
	these characteristics and classified them as active cores if they { were
	detected} in the mid-infrared ($\sim$ 24 $\mu$m) and as quiescent cores
	if no emission at these wavelengths { was measured}. These authors concluded
	that active cores host and form massive stars whereas inactive cores are
	excellent candidates for starless massive cores, prior to the onset of the
	star formation process.

	{ However, unlike low-mass stars ($M <$ 2 M$_{\sun}$), high-mass stars lack a 
	detailed formation scenario. For isolated low mass stars, the shape of the SEDs allows to classify 
	them in four evolutionary classes \citep{1987ARA&A..25...23S,1987IAUS..115....1L,1993ApJ...406..122A}, 
	widely used in the literature. Several factors can be invoked to account for our relatively less 
	detailed knowledge of the formation process/es of high-mass stars (such as: the distances, the 
	relatively small number of massive stars, the amount of energy  and winds emitted by these objects, 
	their short evolutionary time, the fact that they are deeply embedded in the cloud material, etc.). 
	One way to help to improve our understanding of the formation scenario of massive stars is to increase 
	the number of young stars with well-determined parameters and to study their environs.}
	

	\citet{2006A&A...447..221B} cataloged a large number of massive clumps belonging to the southern 
	hemisphere observed in the infrared (IR) continuum at 1.2 mm. From their list, we selected the source 
	IRAS 08589$-$4714 (RA,\,\,DEC(J2000) $=$ 09:00:40.5, -47:25:55) aimed to find massive YSOs (young stellar 
	objects) possibly associated with this source and to analyze their evolutionary stage and derive physical 
	parameters.
	

	IRAS\,08589$-$4714 is located in the giant molecular cloud {\it Vela Molecular Ridge} (see zoomed region 
	in the upper panel of Figure 1), which harbors hundreds of low-mass Class I objects and a large number of 
	young massive stars \citep{1993A&A...275..489L}.  \citet{2006A&A...447..221B} estimated a luminosity
	of 1.8$\times$10$^{3}$ $L_{\sun}$ \citep[compatible with a previous estimate by][]{1989A&AS...80..149W} 
	and a mass of 40 $M_{\sun}$ for this IRAS source. These authors classify the source as an ultracompact 
	HII region (UCHII) since it satisfies  the criteria by \citet{1989ApJ...340..265W}\footnote{These criteria 
	are based on IRAS fluxes satellite of a sample of $\sim$ 1650 UC HII regions previously detected in radio 
	and states that for these regions: $S_{12 \mu m}$ and $S_{25 \mu m}$ $\geq$ 10 Jy,
        $\log(S_{60 \mu m}\,/\,S_{12 \mu m}$) $\geq$ 1.30 and
        $\log(S_{25 \mu m}\,/\,S_{12 \mu m}$) $\geq$ 0.57,
        where $S$ indicates the flux and the subscript the corresponding wavelength.}, although no compact 
        radio-continuum source has been detected \citep{2013A&A...550A..21S}. In spite of its high mass, no CH$_{3}$OH maser 
        emission was found towards this source \citep{1993MNRAS.261..783S}. However, \citet{2013A&A...550A..21S}
        reported water maser emission in the region of the IRAS source.

	\citet{1996A&AS..115...81B} observed the region in the CS(2-1) molecular line. They found that 
	the line has a central velocity $V_{LSR} = +4.3$ km\,s $^{-1}$ and a velocity width at half-maximum
	$\Delta V = 2.0$ km\,s$^{-1}$. In a recent survey, \citet{2014MNRAS.437.1791U} detected emission from
	ammonium molecular tracer, NH$_{3}$, in the (1,1) and (2,2) transitions towards
	the IRAS source. The central velocity coincides with that of the CS line.

	Bearing in mind velocities in the range { 4\,--\,5 km\,s$^{-1}$} for IRAS\,08589$-$4714,  the circular galactic
	rotation model by \citet{1993A&A...275...67B} predicts a kinematical distance of 2.0 kpc, with an 
	uncertainty of 0.5 kpc adopting a velocity dispersion of 2.5 km\,s$^{-1}$ for the interstellar molecular 
	gas.

	\citet{2000A&A...363..744G} and \citet{2008A&A...481..345M} observed this IRAS source in the mid and far 
	infrared and in the sub-millimeter range, and using the IRAS fluxes and those in the J, H, and K bands,
	measured by \citet{1993A&A...275..489L}, they constructed and modeled the SED, and found that the IRAS 
	source has $L \sim 10^{3}$ $L_{\sun}$, masses between 20\,--\,55 $M_{\sun}$, and a B5 spectral type.

	The detection of the IRAS 08589$-$4714 source in the dust continuum, as well as its identification 
	as an UCHII, suggests that it may harbor candidate YSOs embedded. These characteristics make this source 
	interesting to investigate the presence of YSOs and their physical properties. { In 
	\S\,\ref{sec:infrared_sources} we present {\textit{Herschel}} and WISE data and use the WISE color-color 
	diagram to identify YSOs in the region. In \S\,\ref{sec:analysis} we model the corresponding SEDs 
	to derive infalling envelope parameters, such as: mass, size, density and luminosity. Some of the 
	newly detected young stars show an arc-like structure, particular at 12 $\mu$m, pointing to the west, 
	in the direction towards the RCW\,\,38 high star-forming region. 
	In \S\,\ref{sec:arc-like-structures} we suggest that RCW\,\,38 may be contributing to the 
	photodissociation of the IRAS source region, shaping the molecular cloud material around the identified 
	young stars.} Finally, we present the summary and conclusions in \S\,5.


	\section{Infrared sources}
	\label{sec:infrared_sources}

	\subsection{Infrared data}
	\label{subsec:infrared_data}

	The source IRAS\,08589$-$4714 was observed with the {\textit{Herschel}} space
	telescope\footnote{http://www.cosmos.esa.int/web/herschel/}  \citep{2010A&A...518L...1P}
	in the bands of the  Photodetector Array Camera and Spectrometer \citep[PACS, 70 and
	160 $\mu$m;][]{2010A&A...518L...2P} and of the Spectral and Photometric Imaging Receiver
	\citep[SPIRE, 250, 350, and 500 $\mu$m;][]{2010A&A...518L...3G} instruments, with nominal
	{FWHM} of about 5\arcsec, 12$''$, 18$''$, 25$''$, and 36$''$ in the five bands,
	respectively. The data were gathered as part of the {\textit{Herschel}} {\it Infrared Galactic
	Plane Survey} \citep[Hi-GAL,][]{2010PASP..122..314M}, which mapped the Galactic plane in
	mosaics of $\sim 2.3^{\circ} \times 2.3^{\circ}$, taken in the five bands simultaneously.
	IRAS\,08589$-$4714 is located in the HI-GAL field centered at [$\ell,b$] = [268.3$^{\circ}$,
	$-$01.25$^{\circ}$]. These mosaics were obtained in the parallel mode, at a scan
	velocity of 60$''$/s on 2012 November 13. The raw images were reduced to level 1, using
	the HIPE software with standard photometric scripts. The final maps in level 2 were
	processed with the software {\it Scanamorphos}, version 24 \citep{2013PASP..125.1126R}. Finally, 
	we applied appropriate flux correction factors and color corrections to the PACS and SPIRE
	maps, respectively, according to the corresponding manuals\footnote{See Data Analysis Manual
	Guide {\textit{Herschel}} and SPIRE Data Reduction Guide.}.
	
	{To extract the fluxes in each of the {\textit{Herschel}} bands we applied the
	aperture photometry method, using the HIPE software.  We used aperture sizes of 20$''$ and 25$''$
	in the 70 and 160 $\mu$m bands, respectively, and  of $\sim$ 36$''$ in all three SPIRE bands.
	The background was estimated in a ring with internal and external radii of $\sim$ 50$''$ and
	90$''$, respectively, centered on the peak emission position. The photometric uncertainty
	assigned to each flux is given by the standard deviation of the measured values after aperture
	correction. These values are taken at different positions around the central source, without
	background subtraction, using the same aperture as in the flux measurement for the source
	\citep{Exp..Astron..1}. For IRS 1, 4, 5 and 6 fluxes relative errors are
        between 10\,--\,25 \%.  The remaining sources (IRS 2, 3, and 7)
        are contaminated by the emission from the environment in which they are immersed and their fluxes
        for $\lambda >$ 160 $\mu$m have relative errors of $\sim$ 50 \%, preventing us from modeling these 
        sources.}

	This region was also observed in the bands of the {\it Wide-field Infrared Survey Explorer}
	(WISE) satellite\footnote{http: //wise.ssl .berkeley.edu /}  \citep{2010AJ....140.1868W}, at
	W1(3.4 $\mu$m), W2(4.6\,\,$\mu$m), W3(12 $\mu$m), and W4(22 $\mu$m), { with FWHM of about
	6.1$''$, 6.4$''$, 6.5$''$ and 12.0$''$, respectively}. The WISE images and fluxes were obtained 
	from the IRSA ({\it NASA/IPAC Infrared Science Archive}\footnote{http://irsa.ipac.caltech.edu/frontpage/})
	database.

	\subsection{Infrared emission and identification of YSOs}
	\label{subsec:YSOs}

\begin{figure}[!t]
	\includegraphics[width=12cm,height=15cm]{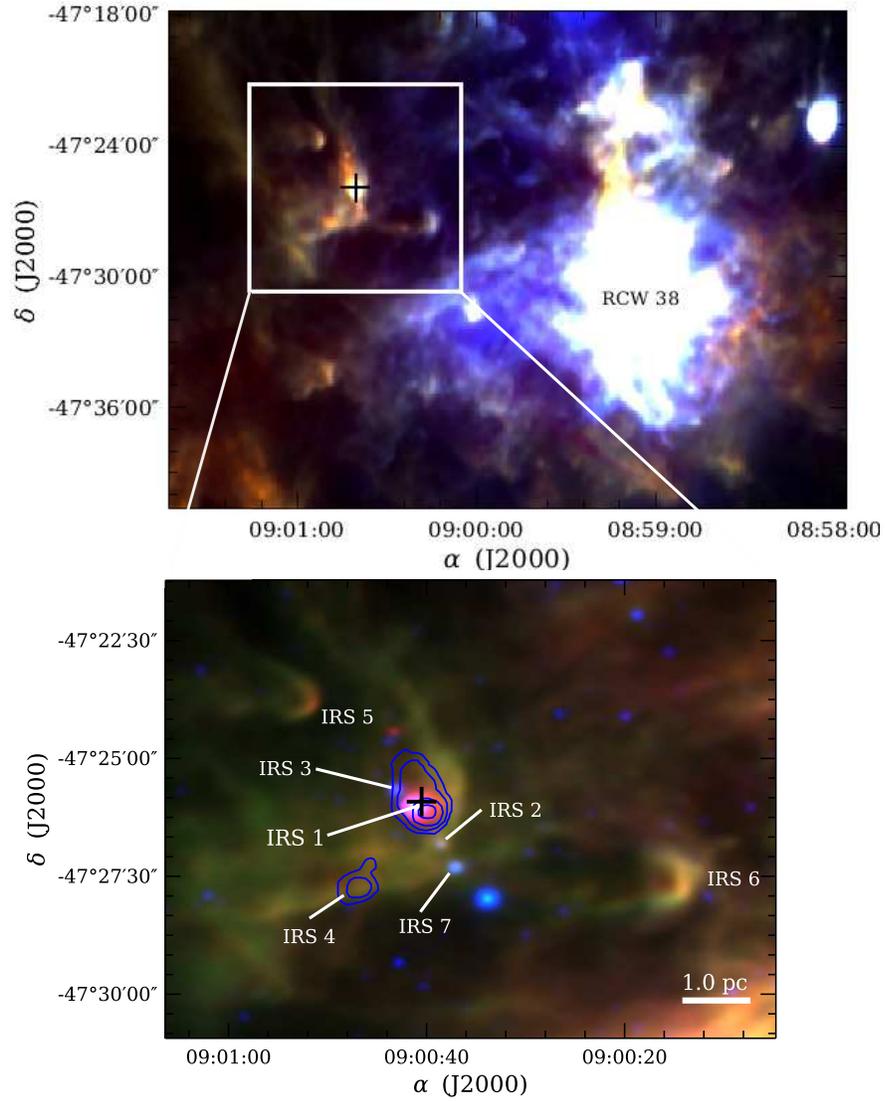}
	\caption{{\it Upper panel:} Composite image showing the emissions at 250 (red), 160 (green)
	and 70 $\mu$m (blue) from {\textit{Herschel}} of the IRAS\,08589$-$4714 region (enclosed by the white rectangle)
        and including the RCW\,38 region. {\it Bottom panel:} Composite image of the WISE images at
        4.6 $\mu$m (blue) and 12 $\mu$m (green), in combination with the {\textit{Herschel}} image at
        70 $\mu$m (red) of the IRAS\,08589$-$4714 region. The $+$ symbol marks the position of the IRAS source. 
        The {\textit{Herschel}} sources labeled from {  IRS} 1 to 7 (see Table \ref{sources}) are also indicated.
        The blue contours show the 1.2 mm dust emission from \citet{2006A&A...447..221B}.} 
	\label{RGB_image}
	\end{figure}

	The bottom panel of Figure~\ref{RGB_image} is a composite image of the WISE images at 
	4.6 $\mu$m (blue) and 12 $\mu$m (green) and the {\textit{Herschel}} image at 70 $\mu$m (red) of the region 
	IRAS\,08589$-$4714. A vast region dominated by the 70 $\mu$m emission is visible to the west. 
	This emission is likely produced by warm dust, while to the east, the emission is dominated by colder 
	dust, detected at $\lambda$\,\,$>$\,\,$100\,\,\mu$m. A transition zone from a hotter to a cooler region is 
	also apparent in this panel (see the highlighted area over the upper panel of Figure \ref{RGB_image}). 
        A well-defined curved edge that crosses the region from north to south, seen in the W3(12\,\,$\mu$m) 
        WISE band (in green) {  in the bottom panel}, is { likely produced} by the emission of warm dust 
        and polycyclic aromatic hydrocarbons \citep[PAHs;][]{2008ARA&A..46..289T}. For more details on this 
        curved and elongated structure see \S\,\ref{sec:arc-like-structures}.

\begin{figure*}
        \includegraphics[width=6.5cm,height=5.5cm]{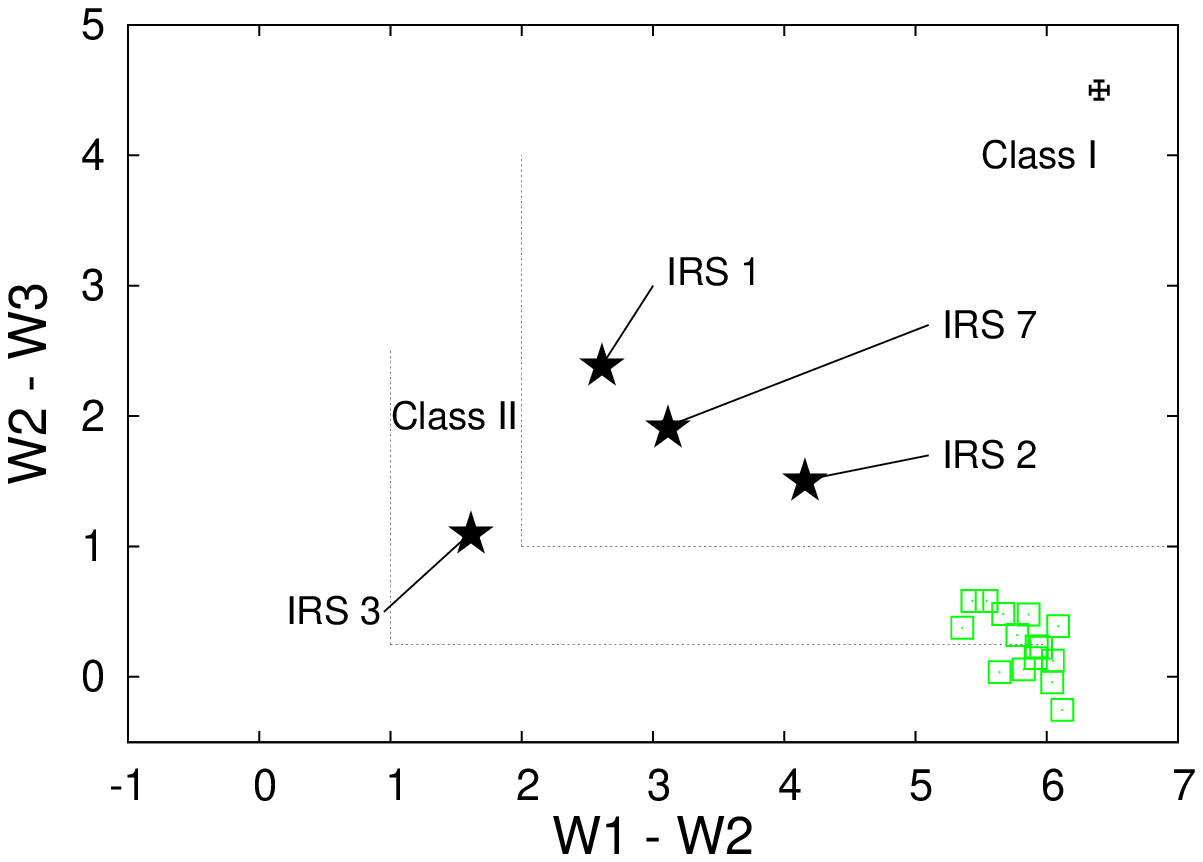}
        \includegraphics[width=6.5cm,height=5.5cm]{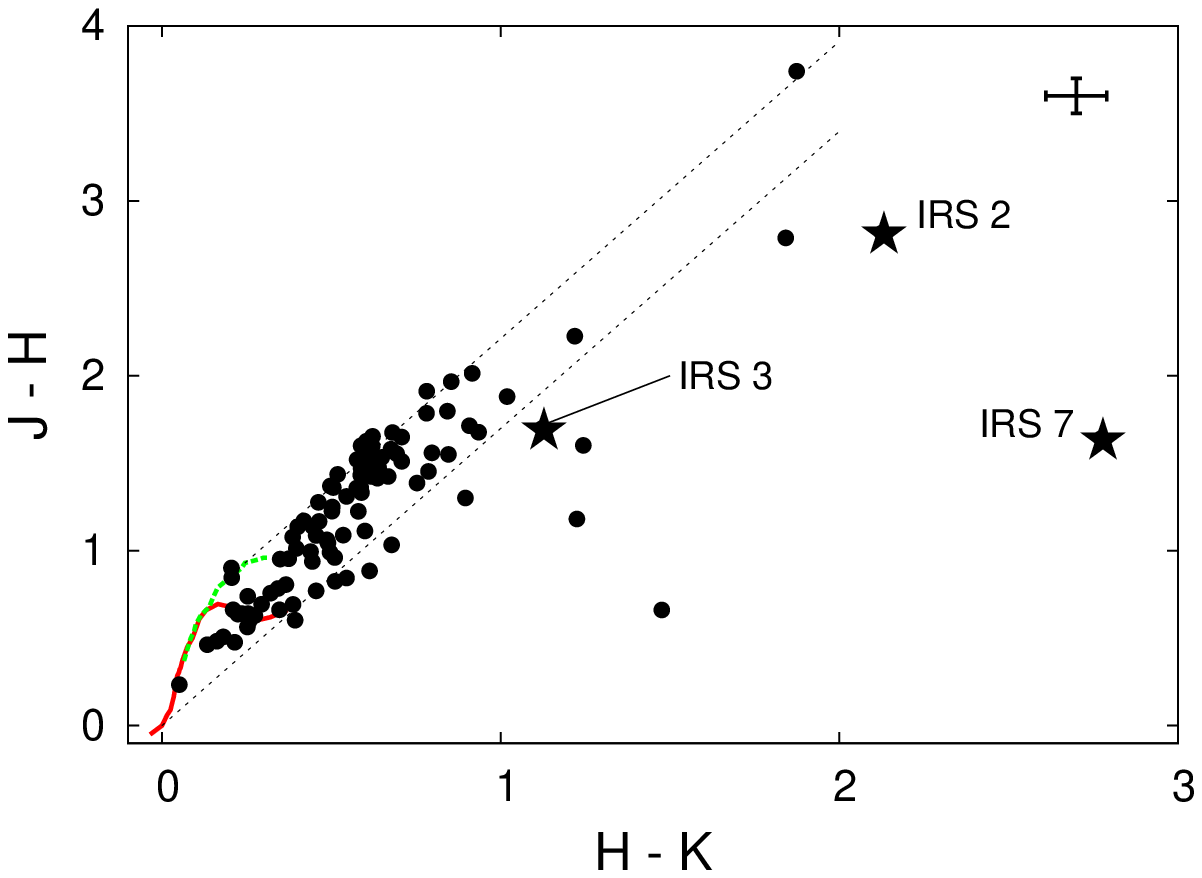}
        \caption{\textit{Left Panel:} WISE W1$-$W2 vs W2$-$W3 color-color diagram. The dotted lines limit the region
         where Class I and Class II objects lie according to the \citet{2012ApJ...744..130K}'s criteria. 
         {  IRS} 1, 2, 3 and 7 are labeled and marked with starred symbols. Green open squares are faint 
         WISE sources likely contaminated by PAH emission \citep{2012ApJ...744..130K}. Average photometric 
         errors are indicated at the upper-right corner. \textit{Right Panel:} 2MASS near-IR J$-$H vs H$-$K 
         color-color diagram. The solid red and green lines mark the
        loci of the main sequence and giant stars \citep{1988PASP..100.1134B}. The dashed lines delineate the
        reddening band for all main-sequence and giant stars \citep{1985ApJ...288..618R}.
        Average photometric errors are indicated at the upper-right corner.}
        \label{cc}
        \end{figure*}

	In the bands of the {\textit{Herschel}} telescope we detected 7 IR sources (IRS, labeled from 1 to 7 in the 
	bottom panel of Figure \ref{RGB_image}), four of which ({  IRS} 1, 2, 3, and, 7) are also detected in
        the bands of the WISE telescope. From the fluxes in these bands, we determine the color indexes
        of the sources. These four WISE detections are potential Class I or II objects, according to the 
        criteria by \citet{2012ApJ...744..130K}\footnote{These criteria state that Class I objects have WISE 
        color indexes satisfying the following inequalities: W1 $-$ W2 $>$ 1.0 and W2 $-$ W3 $>$ 2.0, whereas
	Class II objects have: W1\,\,$-$\,\,W2\,\,$-$\,\,$\sigma_{1}>$ 0.25 and W2 $-$ W3 $-$ $\sigma_{2}>1.0 $, where 
	$\sigma_{1}$ and $\sigma_{2}$ are the combined errors of W1 $-$ W2 and W2 $-$ W3, respectively.}. 
	The left panel of Figure \ref{cc} shows the position of these sources on the WISE color-color diagram. The sources without 
	WISE counterparts ({  IRS} 4, 5, and 6) are probably much younger objects \citep{2009ApJS..181..360C}. 
	IRS 1 and 4 are projected onto two dust clumps (blue contours in Figure\,\ref{RGB_image}) identified by 
	\citet{2006A&A...447..221B} at 1.2 mm. Table \ref{sources} lists the sources detected in the WISE and 
	{\textit{Herschel}} images, with their coordinates and fluxes, and their correlation with the dust 
	clumps detected at 1.2 mm.

\begin{figure*}
        \includegraphics[bb=20 20 650 460,clip,width=\columnwidth]{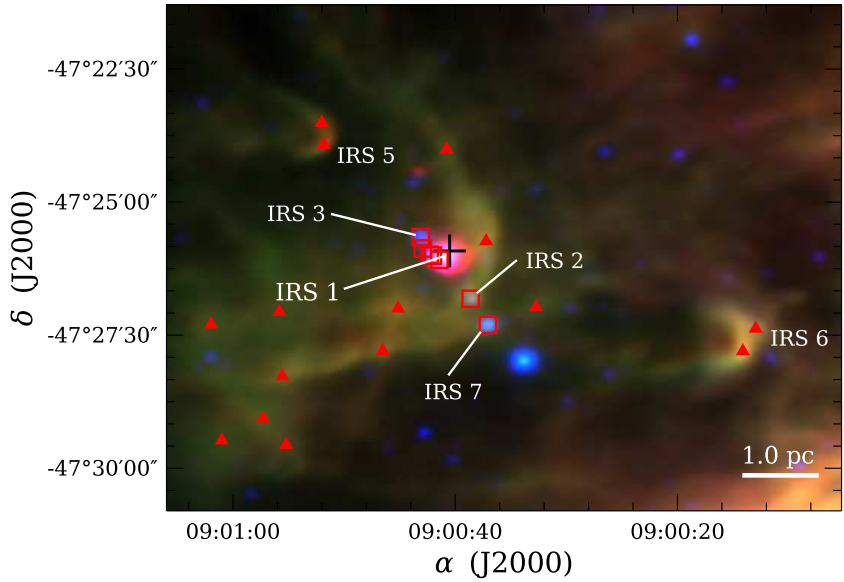}
        \caption{Composite image of the WISE images at 4.6 $\mu$m (blue) and 12 $\mu$m (green), in 
        combination with the {\textit{Herschel}} image at 70 $\mu$m (red) of the IRAS\,08589$-$4714 region. 
        Red triangles show the positions of faint WISE sources listed in Table \ref{sourcesnew}. Red 
        squares correspond to sources with near-IR excesses listed in Table
        \ref{sources_ir}.}
        \label{arcs}
        \end{figure*}

        Green open squares on the left panel of Figure \ref{cc} correspond to faint WISE sources that lie in the region of the 
        WISE color-color diagram contaminated by PAH emission, according to 
        \citet{2012ApJ...744..130K}\footnote{These authors concluded that sources contaminated by PAH emission 
        have WISE colors, such that W1$-$W2 $<$ 1.0 and W2$-$W3 $>$ 4.75.}. PAH emission has prominent lines 
        in the WISE band W1 and W3, producing color excesses \citep{2010AJ....140.1868W}. These sources are 
        listed in Table \ref{sourcesnew}. Interestingly these objects are spatially located around the arc-shape 
        structures near {  IRS} 1, 5 and 6. Figure \ref{arcs} shows the bottom panel of 
        Figure \ref{RGB_image} (4.6 $\mu$m in blue,  12 $\mu$m in green, and 70 $\mu$m in red), indicating with 
        red triangles the positions of these faint WISE sources.  None of these sources has been detected in the 
        {\textit{Herschel}} bands, with exception with those coincident with {  IRS} 5 and 6. All sources in Table 
        \ref{sourcesnew} are potential or candidate YSOs that require deeper observations to confirm their
        nature and evolutionary status. 

        The right panel of Figure \ref{cc} shows the 2MASS near-IR color-color J$-$H vs H$-$K diagram of the IRAS 08589$-$4714
        region. IRS  2, 3 and 7 have near-IR color excesses. Other three sources with near-IR color excess lie in 
        the direction towards IRS\,\,1. All these 2MASS sources, located to the right of the reddening band, are 
        listed in Table \ref{sources_ir}. The 2MASS sources coincident with IRS 5 and 6 do not have color excess in 
        these bands (they are not included in Table \ref{sources_ir}).
        {These sources are marked in Figure \ref{sources_ir} with red squares}. Color excesses in the
        near and middle IR could be associated with warmer dust in the inner part of the envelope which could be 
        due to a circumstellar disk \citep{2006ApJ...638..897S}. This is confirming that these sources are young.
        
	\begin{center}
	\begin{landscape}
	\begin{table}
	\centering
	\setlength{\tabnotewidth}{0.2\columnwidth}
        \tablecols{5}
        \setlength{\tabcolsep}{0.9\tabcolsep}
	\caption{Sources detected towards the IRAS\,08589$-$4714 region using {\textit{Herschel}} images and WISE counterparts}
	\begin{tabular}{c c c D{,}{\,\pm\,}{-1} D{,}{\,\pm\,}{-1} D{,}{\,\pm\,}{-1} D{,}{\,\pm\,}{-1} D{,}{\,\pm\,}{-1} D{,}{\,\pm\,}{-1}}\toprule
	\multicolumn{9}{c}{{\textit{Herschel}}} \\
	Source   &  $\alpha$(2000.0)  &  $\delta$(2000.0)  &  
	\multicolumn{1}{c}{$F_{70}$}  &  
	\multicolumn{1}{c}{$F_{160}$} &
	\multicolumn{1}{c}{$F_{250}$} &  
	\multicolumn{1}{c}{$F_{350}$} &  
	\multicolumn{1}{c}{$F_{500}$} &  
	\multicolumn{1}{c}{1.2\tabnotemark{a} mm}  \\
	  IRS    &    (hh:mm:ss)      &  ($^{o}:\,':\,''$) &    
	  \multicolumn{1}{c}{(Jy)}    &    
	  \multicolumn{1}{c}{(Jy)}    &
	  \multicolumn{1}{c}{(Jy)}    &     
	  \multicolumn{1}{c}{(Jy)}    &    
	  \multicolumn{1}{c}{(Jy)}    &    
	  \multicolumn{1}{c}{(Jy)}     \\ \midrule
	1 & 09:00:40.6 & $-$47:26:01.0 & 236,6   & 343,11 & 201,18 & 88,20 & 25,4    & 1.79,0.17 \\
	2 & 09:00:38.4 & $-$47:26:48.8 & 6.1,2   &  24,6  &  37,12 & 27,8  & 12,3   &  \\
	3 & 09:00:43.1 & $-$47:25:41.5 & 3.4,1   &  29,8  &  46,18 & 40,16 & 19,9    &  \\
	4 & 09:00:48.0 & $-$47:27:31.4 & 7.5,8   &  44,8  &  53,8  & 29,5  & 11,2    & 0.35,0.03 \\
	5 & 09:00:52.0 & $-$47:23:47.1 &11.7,0.7 &  32,4  &  25,3  & 12,1  & 4.9,0.4 &  \\
	6 & 09:00:14.5 & $-$47:27:38.9 &  42,6   &  64,9  &  36,3  & 15,1  & 5.1,0.4 &  \\
	7 & 09:00:36.8 & $-$47:27:17.7 & 5.4,2   &  25,6  &  25,12 & 16,8  &   9,3   &  \\
	\hline
	\multicolumn{9}{c}{WISE} \\
	Source  &  $\alpha$(2000.0)  &  $\delta$(2000.0)  &  
	\multicolumn{1}{c}{W1(3.4 $\mu$m)}  &  
	\multicolumn{1}{c}{W2(4.6 $\mu$m)}  &  
	\multicolumn{1}{c}{W3(12 $\mu$m)}   &  
	\multicolumn{1}{c}{W4(22 $\mu$m)}   &  
	\multicolumn{2}{c}{ID WISE} \\
	  IRS   &     (hh:mm:ss)     &  ($^{o}:\,':\,''$) &     
	  \multicolumn{1}{c}{mag}    &    
	  \multicolumn{1}{c}{mag}    &  
	  \multicolumn{1}{c}{mag}    &  
	  \multicolumn{1}{c}{mag}   &     &  \\ \midrule
	1 & 09:00:40.9 & $-$47:26:01.1 & 9.46,0.03  & 7.07,0.02   & 4.46,0.02 & 0.22,0.02   & \multicolumn{2}{c}{J090040.97$-$472601.1} \\
	2 & 09:00:38.6 & $-$47:26:48.6 & 11.28,0.03 & 9.77,0.03   & 5.61,0.03 & 2.80,0.04   & \multicolumn{2}{c}{J090038.59$-$472648.5}  \\
	3 & 09:00:43.1 & $-$47:25:39.6 & 9.35,0.02  & 8.24,0.02   & 6.63,0.03 & 3.04,0.03   & \multicolumn{2}{c}{J090043.08$-$472539.5}  \\
	7 & 09:00:37.1 & $-$47:27:17.6 & 10.13,0.03 & 8.21,0.02   & 5.10,0.03 & 2.80,0.06   & \multicolumn{2}{c}{J090037.06$-$472718.4} \\ 
	\bottomrule
	\tabnotetext{a}{Beltr\'an et al. (2006).}\\
	\end{tabular}
	\label{sources}
	\end{table}
	\end{landscape}
	\end{center}

\begin{landscape}
\begin{table*}
\centering
\setlength{\tabnotewidth}{0.23\columnwidth}
        \tablecols{3}
  \setlength{\tabcolsep}{1.3\tabcolsep}
\caption{WISE Sources detected towards the IRAS\,08589$-$4714 likely contaminated with PAH emission}
\begin{tabular}{lccccccc} \toprule
ID & $\alpha$(2000.0)  &  $\delta$(2000.0)  &  W1(3.4 $\mu$m)  &  W2(4.6 $\mu$m) &  W3(12 $\mu$m) &  W4(22 $\mu$m) & ID WISE   \\
   &  (hh:mm:ss)   &  ($^{o}:\,':\,''$) &    mag          &    mag         &   mag         &  mag          &   \\ \midrule
WISE 1        & 9:00:12.9 & -47:27:22.5 & 11.34$\,\pm\,$0.04 & 10.76$\,\pm\,$0.03 & 5.22$\,\pm\,$0.02 & 3.34$\,\pm\,$0.05 & J090012.96$-$472722.5\\
WISE 2        & 9:00:14.1 & -47:27:48.0 & 11.19$\,\pm\,$0.03 & 10.96$\,\pm\,$0.06 & 5.03$\,\pm\,$0.02 & 3.20$\,\pm\,$0.07 & J090014.11$-$472748.0\\
WISE 3        & 9:00:45.1 & -47:26:59.9 & 12.92$\,\pm\,$0.08 & 12.53$\,\pm\,$0.08 & 6.44$\,\pm\,$0.05 & 3.25$\,\pm\,$0.03 & J090045.14$-$472659.8\\
WISE 4        & 9:00:57.2 & -47:29:03.7 & 12.72$\,\pm\,$0.06 & 12.59$\,\pm\,$0.09 & 6.55$\,\pm\,$0.05 & 4.64$\,\pm\,$0.06 & J090057.28$-$472903.7\\
WISE 5        & 9:00:32.7 & -47:26:58.4 & 13.09$\,\pm\,$0.08 & 13.03$\,\pm\,$0.07 & 7.21$\,\pm\,$0.11 & 4.80$\,\pm\,$0.06 & J090032.72$-$472658.4\\
WISE 6        & 9:00:46.5 & -47:27:47.5 & 12.25$\,\pm\,$0.05 & 12.29$\,\pm\,$0.07 & 6.25$\,\pm\,$0.02 & 4.19$\,\pm\,$0.04 & J090046.52$-$472747.5\\
WISE 7        & 9:00:55.6 & -47:28:16.3 & 12.79$\,\pm\,$0.12 & 12.21$\,\pm\,$0.11 & 6.78$\,\pm\,$0.15 & 4.29$\,\pm\,$0.14 & J090055.55$-$472816.2\\
WISE 8\tabnotemark{a}  & 9:01:01.0 & -47:29:28.5 & 12.63$\,\pm\,$0.04 & 12.59$\,\pm\,$0.06 & 6.95$\,\pm\,$0.07 & 4.24$\,\pm\,$0.05 & J090101.00$-$472928.4\\
WISE 9\tabnotemark{a}  & 9:00:55.8 & -47:27:03.6 & 13.36$\,\pm\,$0.13 & 12.99$\,\pm\,$0.15 & 7.63$\,\pm\,$0.18 & 5.87$\,\pm\,$0.47 & J090055.83$-$472703.5\\
WISE 10       & 9:01:02.0 & -47:27:17.8 & 13.05$\,\pm\,$0.09 & 12.58$\,\pm\,$0.09 & 6.71$\,\pm\,$0.06 & 4.42$\,\pm\,$0.07 & J090101.97$-$472717.8\\
WISE 11       & 9:00:37.2 & -47:25:43.9 & 11.76$\,\pm\,$0.05 & 11.54$\,\pm\,$0.04 & 5.58$\,\pm\,$0.03 & 3.67$\,\pm\,$0.06 & J090037.21$-$472543.9\\
WISE 12\tabnotemark{a} & 9:00:55.2 & -47:29:33.6 & 12.45$\,\pm\,$0.03 & 12.70$\,\pm\,$0.05 & 6.58$\,\pm\,$0.02 & 4.85$\,\pm\,$0.05 & J090055.21$-$472933.5\\
WISE 13       & 9:00:40.7 & -47:24:00.8 & 12.55$\,\pm\,$0.04 & 12.22$\,\pm\,$0.05 & 6.45$\,\pm\,$0.03 & 4.69$\,\pm\,$0.12 & J090040.76$-$472400.7\\
WISE 14       & 9:00:51.8 & -47:23:55.4 & 12.06$\,\pm\,$0.03 & 11.92$\,\pm\,$0.04 & 6.00$\,\pm\,$0.02 & 3.64$\,\pm\,$0.04 & J090051.82$-$472355.3\\
WISE 15       & 9:00:52.0 & -47:23:30.6 & 12.73$\,\pm\,$0.05 & 12.25$\,\pm\,$0.06 & 6.58$\,\pm\,$0.04 & 4.22$\,\pm\,$0.07 & J090052.00$-$472330.6\\
\bottomrule
\tabnotetext{a}{Source with 2MASS counterpart.}\\
\end{tabular}
\label{sourcesnew}
\end{table*}
\end{landscape}

\begin{landscape}
\begin{table*}
\centering
\centering
\setlength{\tabnotewidth}{0.5\columnwidth}
\tablecols{5}
\setlength{\tabcolsep}{0.9\tabcolsep}
\caption{2MASS sources detected towards IRAS\,08589$-$4714 with near-IR excesses}
\begin{tabular}{lcccccc}\toprule
ID & $\alpha$(2000.0)  &  $\delta$(2000.0)      &  J(1.2 $\mu$m)  &  H(1.6 $\mu$m) &  Ks(2.2 $\mu$m) & ID 2MASS \\
   &  (hh:mm:ss)     &  ($^{o}$:\,$'$:\,$''$) &     mag        &       mag     &   mag     &     \\ \midrule
2MASS 1\tabnotemark{a}   & 9:00:41.4 & -47:26:05.0 & 16.38$\,\pm\,$0.13 & 13.59$\,\pm\,$0.06 & 11.75$\,\pm\,$0.04 & 09004142$-$4726050\\
2MASS 2\tabnotemark{a}   & 9:00:42.0 & -47:26:00.3 &  13.8$\,\pm\,$0.2  &  12.7$\,\pm\,$0.2  & 11.43$\,\pm\,$0.05 & 09004201$-$4726002\\
2MASS 3\tabnotemark{a}   & 9:00:42.3 & -47:25:59.3 & 13.61$\,\pm\,$0.04 & 12.94$\,\pm\,$0.05 &  11.5$\,\pm\,$0.1  & 09004229$-$4725593\\
2MASS 4 (IRS 3) & 9:00:43.1 & -47:25:39.5 & 13.82$\,\pm\,$0.03 & 12.13$\,\pm\,$0.03 & 11.00$\,\pm\,$0.03 & 09004309$-$4725394\\
2MASS 5 (IRS 2) & 9:00:38.6 & -47:26:49.0 &  18.5$\,\pm\,$0.2  & 15.64$\,\pm\,$0.13 & 13.51$\,\pm\,$0.07 & 09003861$-$4726489\\ 
2MASS 6 (IRS 7) & 9:00:37.1 & -47:27:18.4 &  17.4$\,\pm\,$0.2  & 15.77$\,\pm\,$0.15 & 12.99$\,\pm\,$0.03 & 09003706$-$4727184\\
2MASS 7         & 9:00:42.9 & -47:25:52.3 &  16.8$\,\pm\,$0.2  & 15.22$\,\pm\,$0.11 & 13.98$\,\pm\,$0.08 & 09004292$-$4725522\\ \bottomrule
\tabnotetext{a}{Source in the direction towards IRS 1.}\\
\end{tabular}
\label{sources_ir}
\end{table*}
\end{landscape}

	\section{The analysis of the infrared sources}
	\label{sec:analysis}

	\subsection{The spectral energy distributions}
	\label{subsec:YSOs_seds}

        In this Section, we analyze the evolutionary status of some of the sources identified in
        the {\textit{Herschel}} images (see \S\,\ref{subsec:YSOs}) by means of their SEDs. To construct the SEDs 
        we used the fluxes in the {\textit{Herschel}} bands at 70, 160, 250, 350, and 500 $\mu$m. Since protostars 
        lie embedded in dense and opaque material we can derive the main characteristics of the envelopes 
        surrounding the central objects. Fluxes
        for $\lambda <$ 20 $\mu$m were excluded from the modeling of the whole set of sources since
        the code used (DUSTY, see \S\,\ref{subsec:DUSTY_code}) underestimates the emission
        at these wavelengths. This is commonly observed in spherically symmetric models that do not
        include cavities or inhomogeneities in the envelope to allow radiation at these wavelengths
        to escape.

\begin{figure*}[ht]
	\includegraphics[width=6cm,height=5cm]{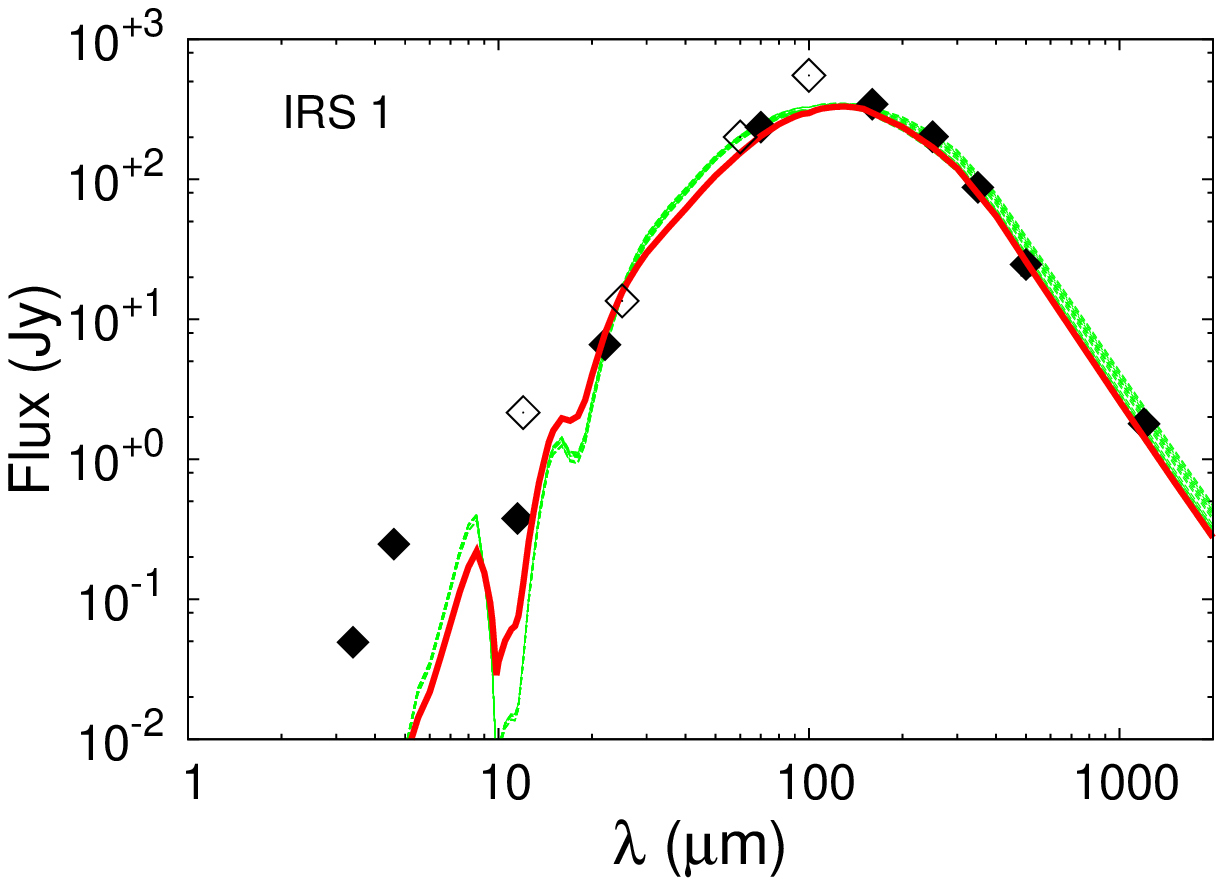}
	\includegraphics[width=6cm,height=5cm]{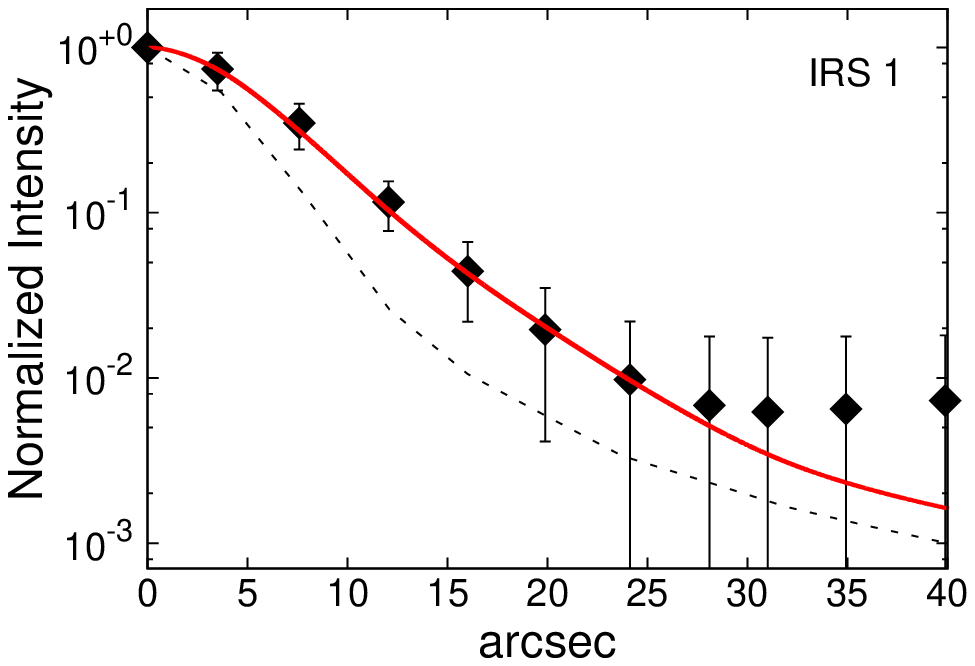}
	\caption{SED (left panel) and intensity profile at 70 $\mu$m (right panel) of {  IRS} 1.
	The filled diamonds correspond to
	the fluxes from the {\textit{Herschel}} and WISE telescopes.  The open diamonds are the IRAS fluxes, shown
	as reference but not fitted. Errors in the fluxes are indicated, except when they are smaller than
	the size of the symbol. The solid red line indicates the best model derived by the DUSTY code.
        The green lines show the 20 DUSTY models comprised within the error bars of the fluxes.
	The dotted line in the right panel is the instrument profile for PACS at 70 $\mu$m
        \citep{2012PICC-ME-TN-033}, used in the convolution with the theoretical intensity profile.}
	\label{sed_psf_f1}
	\end{figure*}
\begin{figure*}[ht]
	\includegraphics[width=6cm,height=5cm]{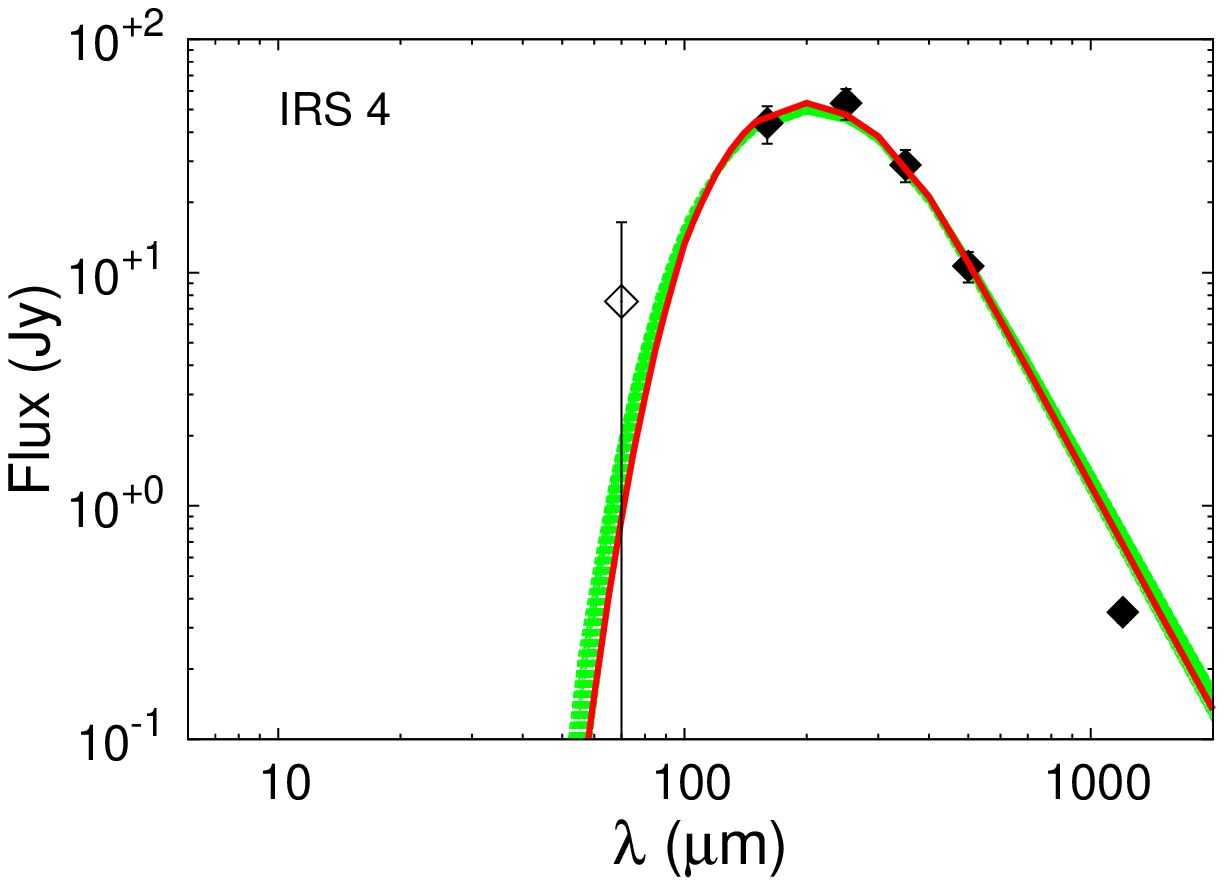}
	\includegraphics[width=6cm,height=5cm]{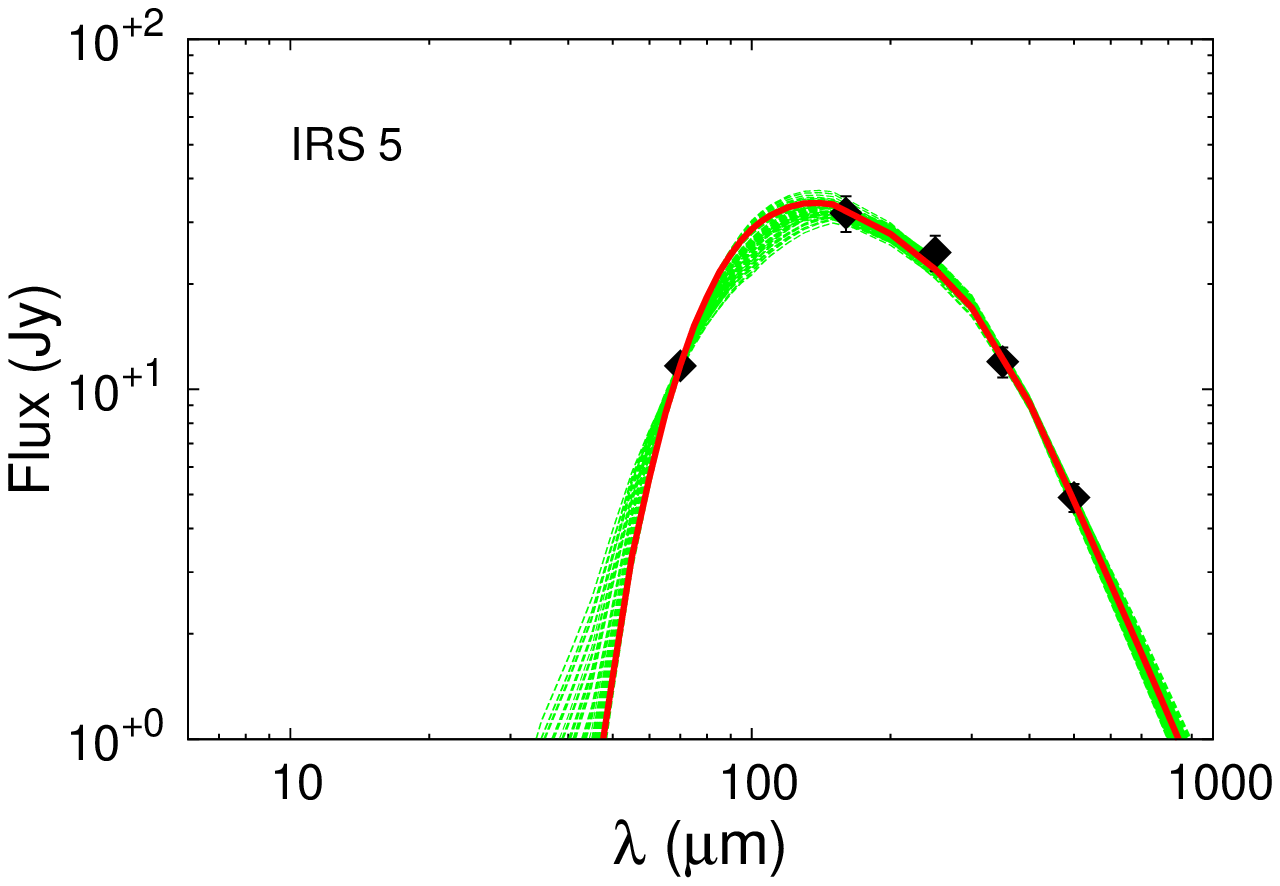}
	\includegraphics[width=6cm,height=5cm]{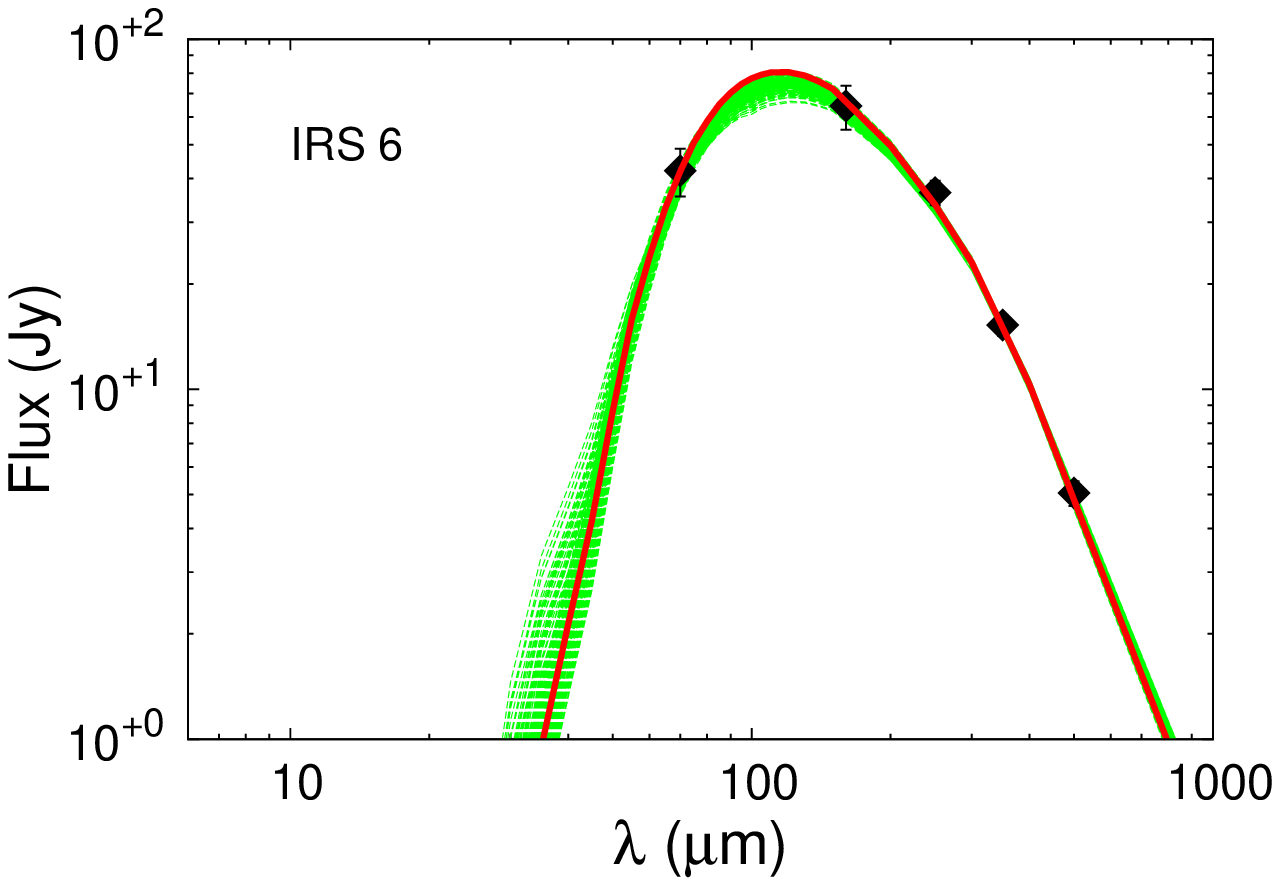}
	\caption{SEDs of {  IRS} 4 (left top panel), 5 (right top panel), and 6 (left bottom panel).
        The filled diamonds belong to the five {\textit{Herschel}} bands.  The empty
	diamond in the SED for {  IRS} 4 is an upper limit to the 70 $\mu$m flux. The error bars are
	included. The relative errors in the fluxes are between 10 and 25\%. The solid red line
        indicates the best model derived by the DUSTY code.  The green lines show 200 DUSTY models
        comprised within the error bars of the fluxes.}
	\label{sed_psf}
	\end{figure*}

        In the case of {  IRS} 1, we model the SED and the 70 $\mu$m radial intensity profile (see Sect. 
        \ref{subsec:70micronprofile}) simultaneously, while for {  IRS} 4, 5, and 6 we model the corresponding SEDs 
        only since they are too faint to obtain reliable profiles at 70 $\mu$m. 
        In the SED corresponding to {  IRS} 1, we include the 1.2 mm flux from
        \citet{2006A&A...447..221B} and the flux at  22 $\mu$m from WISE {  while in the case of the SED for {  IRS} 4  we only 
        include the flux at 1.2 mm. This source does not have a mid-IR counterpart}. The SED of {  IRS} 1 is shown in the 
        left panel of Figure~\ref{sed_psf_f1}, and those of {  IRS} 4, 5, and 6 in Figure~\ref{sed_psf}.

	\subsection{The 70 $\mu$m intensity profile for {  IRS} 1}
	\label{subsec:70micronprofile}

	The average azimuthal intensity profile is commonly employed to compare modeled and observed images.
	This is used to complement the analysis of SEDs, as it allows to reduce the number of free parameters,
	and thus minimizes the degeneration of the solutions \citep{2010A&A...516A.102C}. We present the modeling
	of the intensity profile at 70 $\mu$m corresponding to {  IRS} 1.

	{ Since the PSF at 70 $\mu$m is rather elongated \citep[due to the scan speed, which for 60$''$/s 
	gives a PSF of 5.83$''\times$12.12$''$,][]{2010A&A...518L...2P}, to determine the radial intensity profile,  
	several one-dimension radial profiles are obtained in different directions from the center of the 
	source. In the case of IRS 1 (see Figure\,\,\ref{sed_psf_f1}), we used only the hemisphere in the opposite 
	direction to the apex of the curved structure seen in Figure \ref{RGB_image} to avoid contamination 
	from the surrounding diffuse and extended emissions. The uncertainty in the profile is given by the 
	noise in the image and the non-circularity of the beam of the source. The latter effect is taken into 
	account by the standard deviation of the average azimuthal flux. The profile is normalized by its 
	maximum value.}


	\subsection{The fitting method}
	\label{subsec:DUSTY_code}

	The SEDs of the four sources, as well as the 70 $\mu$m intensity profile of IRS\,\,1, were modeled
	with the DUSTY\footnote{http://www.pa.uky.edu/$\sim$moshe/dusty/} code of \citet{1997MNRAS.287..799I}. 
	The code solves the radiative transfer in one-dimension for a dusty shell that surrounds a central source,
	which emits as a black body, whose radiation is absorbed, scattered and re-emitted by the dust in
	the envelope. This envelope has a density profile that follows a power law ($ n \propto r ^{-p}$),
	and it is parameterized by the relative size given by $Y = r_{\rm ext}/r_{\rm in}$, where
	$r_ {\rm ext}$ and $r_ {\rm in}$ are the external and internal radii of the shell. The temperature
	of the central source ($T_ {\rm star}$) and the temperature at the internal radius of the envelope
	($T_{\rm in}$) are fixed, with values of 15\,000 and 300 K, respectively. { The results are rather
	insensitive to the values of $T_{\rm star}$. 
	
	We tested the DUSTY SEDs for T$_{star}$ between 5\,000 and 
	50\,000\,\,K, and no distinguishable differences in the results were found.} 
	{ On the other hand, the inner radius is relatively unconstrained by current data resolution 
	(5$''$ corresponds to $\sim$ 10\,000 UA at 2 Kpc), as is T$_{in}$. As a matter of fact, T$_{in}$ determines 
	the radius at which the code starts the calculations. Several SEDs modeling, as well as mid-infrared 
	spectra, of massive stars rule out inner radius temperatures consistent with dust sublimation temperatures 
	($\sim$ 1500 K). The radiation pressure, stellar winds and$/$or shocks are usually invoked to produce a 
	large cavity depleted of dust and to limit the amount of short wavelengths emission 
	\citep{1990ApJ...354..247C,1998ApJ...500..280F,2000A&A...357..637H}. For T$_{in}$ we fixed a value in a 
	similar manner as previous works in the literature, giving an inner radius 
	larger than it would be expected from dust sublimation \citep{2000A&A...357..637H,2002A&A...389..908J,
	2009A&A...506.1229C,2010A&A...516A.102C,2010A&A...520A..78V,2012ApJ...757..113H}.} Finally, for very young 
	sources it is usual to assume that the envelope is composed of particles surrounded by a thin layer of ice. 
	For this reason, we adopt the opacity corresponding to a density of $10^{6}$\,\,cm$^{-3}$ from 
	\citet{1994A&A...291..943O}.

	The DUSTY model needs to be scaled to the distance and the bolometric luminosity ($L_ {bol}$) of
	the central source to be able to compare the models and observed data. The bolometric luminosity
	is, however, an output parameter, since it is calculated by integrating the modeled SED.
	Consequently, the luminosity is {  re-evaluated as input parameter} iteratively {  until minimizing} the difference between the
	models and observations. We adopted a distance of 2.0 kpc (see \S\,\ref{sec:intro}), assuming that
	all the sources belong to the IRAS\,08589$-$4714 region.

	On the other hand, the intensity profiles modeled by DUSTY show an intense and very narrow central component
	(width $<$ 1\arcsec\ ). This component is the stellar radiation attenuated by a dusty envelope,
	whose width is proportional to the stellar radius \citep{1996MNRAS.279.1011I}. { Given that the pixel
	size at 70 $\mu$m is 3\farcs 2, 
	we re-binned the model profile using a slightly larger bin of 4$''$}. 
	{Finally, this profile is convolved with the instrumental profile at 70\,\,$\mu$m, taken
	as the average intensity profile of PACS calibrated sources, such as: $\alpha$\,Tau, Red Rectangle, IK Tau and 
	the Vesta asteroid \citep{2012PICC-ME-TN-033}. The first two sources are used to model the core and the 
	last two the extended wings of the profile \citep{2011PASP..123.1218A}.}

	To systematically compare the models with the observations we built a grid of 127\,100 DUSTY models.
	The grid was constructed for 31 values of the power index ($p$) in the density profile, ranging
	from $0.0$ to $3.0$ in steps of 0.1, 82 values of the envelope relative size ($Y$) in the range
	100-910 with steps of 10, and 50 values of the optical depth ($\tau_{\rm 100}$), from 0.1 to 5.0, in steps of 0.1.
	The space of the parameters explored is similar to that used by \citet{2010A&A...516A.102C},
	restricted to a more limited range of each parameter but including typical values for young stellar objects. Table
	\ref{set_param} summarizes the ranges of the free parameters.  The best fit is obtained by means
        of the goodness-of-fit criterion. We employed the weighted
        minimum square method to fit the SED.  The weights are inversely proportional to the
        square of the errors assigned to each observed data point.

	\begin{table}
	\centering
	\caption{Range explored for the different parameters with the DUSTY code}
	\begin{tabular}{ccc}\toprule
	Parameter    &     Range      &  Step \\ \midrule
	$p$           &  0.0\,--\,3.0    &  0.1  \\
	$Y$           &  100\,--\,910    &  10  \\
	$\tau_{100}$  &  0.1\,--\,5.0    &  0.1  \\
	$T_{star}$    &  15\,000 K      &  $-$  \\
	$T_{in}$      &  300 K          &  $-$  \\ \bottomrule
	\hline
	\end{tabular}
	\label{set_param}
	\end{table}

	\begin{table}
	\centering
	\caption{Best Modeled parameters}
	\begin{tabular}{cccccc} \toprule
	\multicolumn{6}{c}{Input model parameters} \\
	\hline
	IRS              &     $L$       &  $\tau_{100}$ &  $p$ &  $Y$\\
	&   ($L_{\sun}$) &           &      &     \\ \midrule
	1  &    1.9$\,\times 10^{3}$ &  0.2  &  0.0 & 250  \\
	4  &    1.2$\,\times 10^{2}$ &  4.8  &  0.4 & 130  \\
	5  &    1.2$\,\times 10^{2}$ &  3.9  &  1.7 & 410  \\
	6  &    3.0$\,\times 10^{2}$ &  5.0  &  2.1 & 110  \\ \bottomrule
	\multicolumn{6}{c}{Derived physical parameters} \\
	\hline
	IRS         &  $r_{in}$ &  $r_{ext}$ &  $T_{dust}$ &  $M_{env}$  &  $n(r_{in})$         \\
	&  (AU)     &  (pc)      &    (K)     &  ($M_{\sun}$) &  (cm$^{-3}$) \\ \midrule
	1  & 112  &  0.14  & 18   & 86  &  3$\,\times 10^{5}$ \\
	4  &  36  &  0.02  & 16   & 42  &  2$\,\times 10^{7}$ \\
	5  &  66  &  0.13  & 10   & 38  &  1$\,\times 10^{7}$ \\
	6  & 129  &  0.07  & 15   & 11  &  7$\,\times 10^{6}$ \\ \bottomrule
	\end{tabular}
	\label{param}
	\end{table}
	
	\begin{table}
	\centering
	\setlength{\tabnotewidth}{12cm}
	\tablecols{6}
	\setlength{\tabcolsep}{1.0\tabcolsep}
	\caption{Average modeled parameters}
	\begin{tabular}{c D{,}{\,\pm\,}{-1} D{,}{\,\pm\,}{-1} c D{,}{\,\pm\,}{-1} c} \toprule
	\multicolumn{6}{c}{Average input parameters} \\
	\midrule
	   &  \multicolumn{1}{c}{IRS}   &  \multicolumn{1}{c}{$\tau_{100}$} &  $p$ &  \multicolumn{1}{c}{$Y$} \\
	                      &               &      &     \\
	\hline
	  &  1  &  0.3,0.0\tabnotemark{*}  &  0.8$\,\pm\,$0.1 & 330,53  \\
	  &  4  &  3.3,1.0        &  0.4$\,\pm\,$0.3 & 214,56  \\
	  &  5  &  3.0,1.5        &  1.4$\,\pm\,$0.4 & 447,176  \\
	  &  6  &  2.7,1.5        &  1.5$\,\pm\,$0.6 & 170,62  \\
	\bottomrule
	\multicolumn{6}{c}{Average derived physical parameters} \\
	\hline
	IRS         &  \multicolumn{1}{c}{$r_{in}$} &  \multicolumn{1}{c}{$r_{ext}$} &  $T_{dust}$ &  \multicolumn{1}{c}{$M_{env}$}  &  $n(r_{in})$         \\
	            &  \multicolumn{1}{c}{(AU)}     &  \multicolumn{1}{c}{(pc)}      &    (K)     &  \multicolumn{1}{c}{($M_{\sun}$)} &  (cm$^{-3}$) \\
	\midrule
	1  & 129,2  &  0.21,0.03  & 16$\,\pm\,$1   & 124,33  &  (4.00$\,\pm\,$0.02)$\times 10^{5}$ \\
	4  &  35,5  &  0.04,0.01  & 14$\,\pm\,$2   &  60,12  &  (1.6$\,\pm\,$0.4) $\times 10^{7}$ \\
	5  &  56,16 &  0.13,0.08  & 11$\,\pm\,$2   &  38,14  &  (8.4$\,\pm\,$2.5) $\times 10^{6}$ \\
	6  &  89,29 &  0.08,0.05  & 15$\,\pm\,$3   &  15,3   &  (4.9$\,\pm\,$1.3) $\times 10^{6}$ \\
	\bottomrule
	\tabnotetext{*}{All models that fall within the fluxes error bars have the same $\tau_{100}$ value.}
	\end{tabular}
	\label{average_param}
	\end{table}

	In the case of {  IRS} 1, to find a model that fits both the SED and the intensity profile
	simultaneously, we follow the procedure used by \citet{2010A&A...516A.102C}. We first fixed the
	value of the opacity ($\tau_{\rm 100}$) and by fitting the intensity profile we obtained an
	initial value for the power law density index ($p$) and for the envelope relative size ($Y$).
	Then we recalculated the value of $\tau_{\rm 100}$ by fitting the SED, using the values of
	$p$ and $Y$ derived in the previous step. We adopted the new value of $\tau_{\rm 100}$ in the
	next iteration and the process was repeated. The convergence was achieved in a few steps.
	This procedure considers the fact that the intensity profile is strongly dependent on the size
	of the envelope and the density distribution, whereas the SED depends mainly on the column density,
	that, in turn, depends on the opacity \citep{2010A&A...516A.102C}.


	\subsection{Derived parameters}
	\label{subsec:parameters}

	The solid red line in Figs.~\ref{sed_psf_f1} and \ref{sed_psf} shows the best DUSTY models for
	{  IRS} 1, 4, 5, and 6. {With green continuous line we show all DUSTY models comprised within 
	the error bars of the fluxes (20 in the case of IRS 1 and 200 for IRS 4, 5 and 6).} Table \ref{param} 
	summarizes the parameters that best reproduce the data: the luminosity $L$, the opacity $\tau_{\rm 100}$, 
	the power law density index $p$, and the relative size of the envelope $Y$. We include in this table 
	the physical parameters of the envelope: the inner ($r_{in}$) and external ($r_{ext}$) radii, the 
	temperature of the dust at $r_{ext}$ ($T_ {dust} $), and the mass ($M_{env}$) and the density in the 
	inner part of the envelope ($n(r_{in})$).

	{ Since the Dusty code does not provide error determinations for the modeled parameters,
	Table \ref{average_param} lists average values and standard deviations corresponding to all 
	DUSTY models, calculated for the same luminosity, that fall within the error bars of the fluxes 
	(green lines in Figs. \ref{sed_psf_f1} and \ref{sed_psf}). Best model parameters listed in Table 
	\ref{param} are not always within the standard deviations given in Table \ref{average_param}. 
	However, they are not significantly different from the average values. In the case 
	of IRS 1, $p$ and $Y$ are derived modeling simultaneously the SED and the 70 $\mu$m intensity profile. 
	For this reason, the differences between the values listed in Table \ref{param} and 
	\ref{average_param} are larger.}
	
	For the SED of {  IRS} 1, the fluxes at 3.4, 4.6, and 12 $\mu$m are not well fitted by the
	model, as we have mentioned before. These fluxes are likely contaminated by shocks and PAH
	emission features detected in these WISE bands \citep{2010AJ....140.1868W}. {Moreover, \citet{Saldanio-2016} detected an  
	outflow associated with IRS 1, which likely produces a cavity in the envelope. 
	{ Dust in the walls of the cavity is heated by UV and optical radiation coming from the stellar 
	embryo/s, re-irradiating in the IR \citep[see, for example,][]{2009ApJ...700..872B,2013ApJ...766...86Z}. 
	Thus mid-IR fluxes are considered as upper limits.}
	In addition, the observed 
	intensity profile is {  well fitted by 	the modeled profile { out to} 20$''$, but not beyond.}
	
	The density profile for the envelope of {  IRS} 1 derived with the DUSTY code is constant,
	and consistent with the supposition introduced by \citet{2000A&A...363..744G} in their modeling of the
	SED of this source. 
	In addition, the other parameters determined by these authors 
	($M =55\, \, M_{\sun} $, $R_{\rm max}=0.2$ pc, $L$\,\,=\,\,2.4\,\,$\times 10^{3}$ $L_{\sun}$) are similar
	or the same order as the parameters obtained through DUSTY. However, we derived an opacity
	for the envelope four times larger than the value obtained by \citet{2000A&A...363..744G}. The difference 
	is probably due to the fact that these authors used silicate and graphite dust grains without
	a layer of ice on the surface, producing less attenuation to radiation. The inclusion of a
	layer of ices gives a more realistic model \citep{1994A&A...291..943O}.

	\subsection{Virial Mass}
	{  Assuming that: 1) the parent 
	cloud is similar to that one used in the DUSTY code; 2) the source is in hydrostatic equilibrium; 
	and 3) the volume density of the gas decays as a power law ($\rho \propto r^{-p}$), we estimate 
	the envelope mass using the following expression for the virial theorem:  
	\begin{equation}
	\label{virial_mass}
	M_{vir} = 3\left(\frac{5-2p}{3-p}\right)\frac{r_{ext}}{G\,k} \, \sigma^{2}, 
	\end{equation}

	\noindent
	where $G$ is the gravitational constant and $\sigma$ is the one-dimensional velocity dispersion averaged 
	over the entire system. The $k$ factor is approximately { equal to} 1 for $Y \gg 1$ and $p <$ 2.5. Using the 
	parameters calculated by the DUSTY code, such as $p$ and $r_{ext}$, and $\Delta v\,=\,\sqrt{8\,ln(2)}\,\sigma$,
	where $\Delta v$ is the full-width at half-maximum (FWHM) of the
	C$^{18}$O line detected towards IRS\,\,1 \citep{Saldanio-2016}, we estimate $\sigma =$ 0.65$\,\pm\,$0.05\,km\,s$^{-1}$
	and a virial mass for {  IRS} 1 listed in Table \ref{param_virial}. For the other three starless cores 
	(IRS 4, 5 and 6), we adopt an average velocity dispersion $\sim$ 0.55$\,\pm\,$0.25\,km\,s$^{-1}$,
	averaging the values determined by \citet{2013MNRAS.432.3288S} and \citet{2014ApJ...787..113B} for multiple starless cores
	observed in NH$_{3}$ molecular transitions. Table \ref{param_virial} lists the calculated values.
	Sources 1, 5 and 6 have virial masses roughly in agreement with their DUSTY mass envelope, confirming that 
	they are gravitationally bound. On the other hand, {  IRS} 4 has a virial mass smaller than the envelope mass, indicating
	that this source may be collapsing.}
	
	\begin{table}
	\centering
	\caption{Masses for four infrared detected sources}
	\begin{tabular}{lcccc}
	\toprule
	        &    IRS 1  &   IRS 4   &   IRS 5   &    IRS 6  \\ 
	        &      $M$ ($M_{\sun}$)     &     $M$ ($M_{\sun}$)     &      $M$ ($M_{\sun}$)    &      $M$ ($M_{\sun}$)     \\
	\midrule
	Virial Mass       &  67  &   8  &  34  & 14 \\
	Envelope Mass     &  86  &  42  &  38  & 11  \\
	\bottomrule
	\end{tabular}
	\label{param_virial}
	\end{table}
%
	\section{Probable scenario of the origin of the arc-like structures}
	\label{sec:arc-like-structures}

	In Figure~\ref{mapa_continuo} we show large scale WISE and {\textit{Herschel}} images of the
	region of the IRAS 08589$-$4714 source with the aim of explaining the arc-like
	structures detected mainly at 12 $\mu$m. We superimpose contour lines of the
	emissions at 12 $\mu$m from WISE (green lines) and at 1.2 mm (blue lines) from
	\citet{2006A&A...447..221B}. The seven sources identified and labeled in Figure~\ref{RGB_image}
	are also marked onto the 4.6 $\mu$m image.

	\begin{figure*}
	\centering
	\includegraphics[bb=20 16 625 410,clip,width=11.8cm,height=7.5cm]{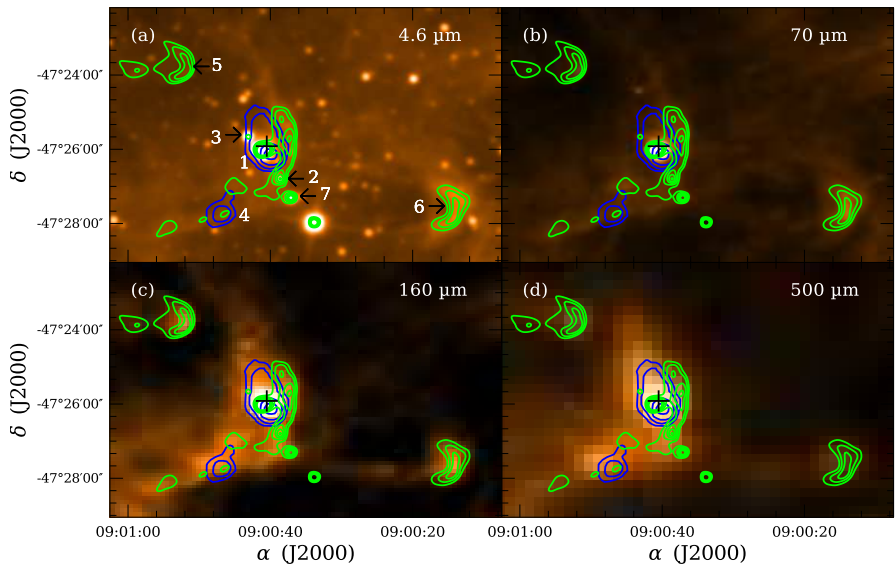} 
	\caption{WISE 4.6 $\mu$m and {\textit{Herschel}} 70, 160 and 500 $\mu$m images of the IRAS\,\,08589$-$4714 region.
	Blue contours correspond to the 1.2 mm emission \citep{2006A&A...447..221B} and those in green to the WISE
	12 $\mu$m emission. The sources labeled from {  IRS} 1 to 7 correspond to those in Table \ref{sources}.}
	\label{mapa_continuo}
	\end{figure*}

	At 12 $\mu$m we can see three arc-like structures pointing to the west. The more
	extended structure borders the western rim of {  IRS} 1, and the two smaller ones
	coincide with {  IRS} 5 and 6. 
        \citet{Saldanio-2016} observed $^{12}$CO($3-$2),
        $^{13}$CO($3-$2), C$^{18}$O($3-$2), HCO$^{+}$($3-$2), and HCN($3-$2) molecular lines in
        a region of 150$''$ $\times$ 150$''$, centered on the IRAS source with the APEX telescope. These data covered
        {  IRS} 1 and 3 positions.  The molecular emissions of $^{12}$CO and $^{13}$CO
        to the west of {  IRS} 1 are very weak. In addition, the integrated emission of $^{13}$CO
	shows a very steep intensity  gradient to the west, suggesting that the material linked to
        {  IRS} 1 is being compressed. Finally, in
	Figure~\ref{mapa_continuo} we can see that the emission at 12 $\mu$m  borders the
	western rim of the cold dust emission, as it is better seen in the 160 and 500 $\mu$m
	images. The WISE filter at 12 $\mu$m (with a bandwidth $\sim$ 9 $\mu$m) includes
	strong spectral lines of neutral and ionized PAH \citep{2008ARA&A..46..289T}, typical of
	photodissociation regions (PDRs). The comparison of the emission at 12 $\mu$m likely from
	ionized PAHs and at 1.2\,\,mm from cold dust suggests the existence of hot sources located to
	the west, which are creating a PDRs, bordering the molecular gas.

	We have used the 70 and 160 $\mu$m {\textit{Herschel}} images to obtain a dust temperature
	map { at the highest angular resolution possible in a relatively small area. In addition, these 
	bands are good tracers of warm dust, a likely scenario for IRAS 08589$-$4714 since not only is exposed 
	to an intense external radiation field but also contains YSO candidates embedded.}	
	The map was constructed as 
	the inverse function of the ratio of {\textit{Herschel}} 70 and 160 $\mu$m maps, i.e., $T_{c} = f(T)^{-1}$.
	Assuming a dust emissivity following a power law \hbox{$\kappa_{\nu}$ $\propto$ $\nu^{\beta}$},
	being ${\beta}$ the spectral index of the thermal dust emission,   in the optically thin thermal
	dust emission regime $f{(T)}$ has the form:

	\begin{equation}
	\qquad f{(T)} = \frac{S_{70}}{S_{160}} = \frac{B(70,T)}{B(160,T)} \left( \frac{70}{160} \right) ^{\beta}
	\end{equation}

	\noindent
	where $B(70,T)$ and $B(160,T)$   are the blackbody Planck function for a temperature $T$ at
	the wavelengths 70 $\mu$m and 160 $\mu$m, respectively. The pixel-to-pixel temperature was
	calculated assuming a typical  value  $\beta$ = 2. 

        In Figure~\ref{fig:tdust} we show the
	color-temperature map obtained using the method explained above. For temperatures
        in the range $\sim$ 20--40 K, the uncertainty in derived dust temperatures using this
        method was estimated to be about $\sim$ 10--25 $\%$.  
        Also, Figure~\ref{RGB_image} shows that the interstellar
        dust to the west of the sources appears warmer than to the east because of the presence
        of emission at lower wavelengths. 
        Figure \ref{fig:tdust} suggests the existence of a gradient in the dust temperature, with the lower values
        to the east. Note in particular the regions to the east of {  IRS} 5 and 6, where dust appears to
        be shielded from UV photons. We caution that $\sim$ 40$\%$ of the area shown
        in Figure~\ref{fig:tdust} has temperatures $<$\,\,20\,\,K where, as mentioned, uncertainties may be
        large. However, any overestimation makes the gradient less { steeper}. In other words, real temperatures
        below 20 K would be even lower than shown in this figure and the gradient { steeper}. Thus, the gradient
        from west to east, exists in spite of the uncertainties in T. { However $\sim$ 60\% of the area shown in 
        Figure\,\,\ref{fig:tdust} is dominated by a relatively warmer component that prevails in the emission at 
        70 $\mu$m towards the west, which likely indicates the influence of RCW 38 (see below).}

	\begin{figure}[ht!]
	\centering
	\includegraphics[width=275pt]{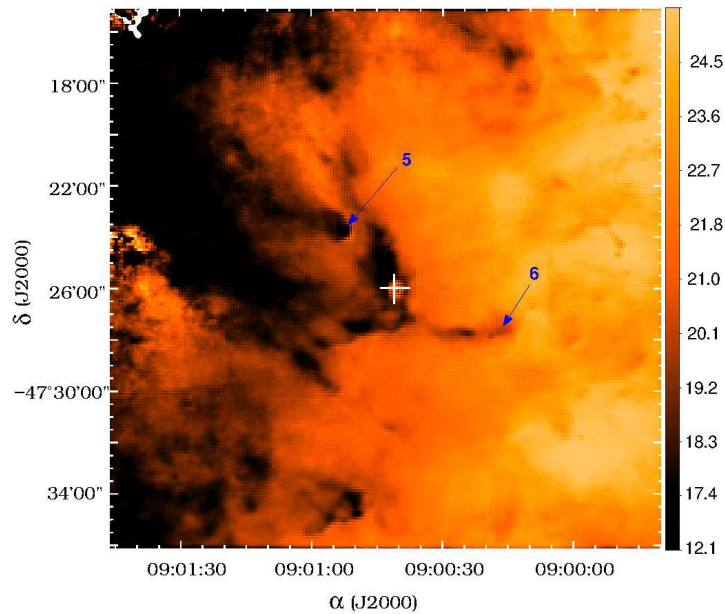}
	\caption{Dust  temperature map  (in color scale) derived from {\it Herschel} emission at 70
	and 160 $\mu$m. The color-temperature scale (in K) is on the right. The white cross indicates the IRAS
	source position. The arrows mark the positions of {  IRS} 5 and 6, as references.}
	\label{fig:tdust}
	\end{figure}

        About 16\farcm 7 to the west of IRAS 08589$-$4714 lies the high mass star-forming
        region RCW 38 (upper panel of Figure~\ref{RGB_image}). This HII region is excited by an O5.5 binary star
        \citep[RCW\,38 IRS 2;][]{2009AJ....138...33D}. In addition, \citet{2011ApJ...743..166W}
        reported the detection of a dozen O-type stars and several candidates OB stars associated
        with RCW\,38. Many of these sources are ionizing the interstellar medium, generating HII
        regions located in the surrounding area
        \citep{1997ApJ...488..224K,1999PASJ...51..791Y,2003A&A...397..213P,2014yCat..22110029B,
        2014yCat..22120001A}.  Figure \ref{fig:wcr38} { indicates} the location of these HII regions. 
        Moreover, the center of RCW\,38 shows strong X-ray activity that may be
        produced by the formation of massive stars. There are, at least, 50 sources emitting at these
        wavelengths \citep{2013ApJS..209...27K,2013ApJS..209...32B}. RCW\,38 is located at a distance
        of 1.7 kpc \citep{2009MNRAS.400..731S}, similar, within errors, to the distance of IRAS 08589$-$4714
        (2.0 $\pm$ 0.5 kpc).  The center of the RCW\,38 lies $\sim$ 10 pc away from the center of the
        IRAS 08589$-$4714 region. 

        \begin{figure}[ht!]
        \centering
        \includegraphics[bb=18 11 590 460,clip,width=300pt]{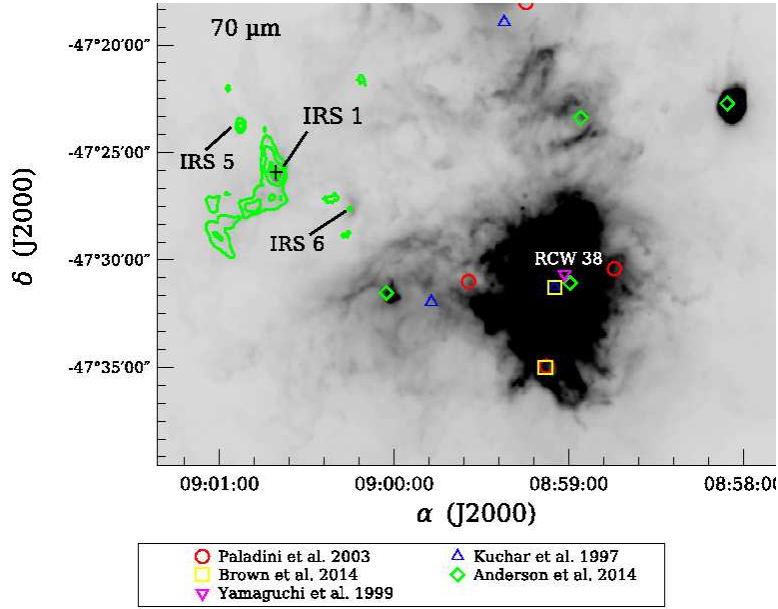}
        \caption{70 $\mu$m {\textit{Herschel}} images of the WCR 38 region showing the locations of different
         HII regions. References are included at the bottom of the figures. 
        The cross shows the position of IRAS 08589$-$4714 source, on top of which 70 $\mu$m
        contours are superimposed.} 
        \label{fig:wcr38}
        \end{figure}

\citet{2013A&A...556A..92K} carried out a large-scale mapping of RCW\,38 in the [CII] and PAH emissions.
They found that [CII] (158 $\mu$m) emission extends about 10$'$ (about 5 pc) away from
the center of  RCW\,38, in particular in the north and east directions.
Sources IRS 1, 2, 7 and 6 lie about 2.5$'$ or about 1.5 pc away from
the most external [CII] contour. It is noteworthy that [CII] is considered one of the main cooling
agents in low-density photodissociation regions with PAHs emissions
\citep[see, for example,][]{1999RvMP...71..173H, 2003ANS...324..139G} and,
therefore, a good tracer of them. In addition, [CII] emission strongly correlates with PAHs
emissions.  Consequently, it is likely that massive stars in RCW\,38 are photodissociating the molecular gas producing the PDR
and heating the interstellar dust to the west of IRAS\,\,08589$-$4714.

\section{Summary and conclusions}
\label{sec:summary}

We use WISE 3.4, 4.6, 12 and 22 $\mu$m and {\textit{Herschel}} 70, 160, 250, 350 and 500 $\mu$m fluxes to
analyze 7 sources identified in the IRAS 08589$-$4714 region. 
{ Four of these sources (called IRS 1, 2, 3 and 7) have WISE colors of Class I and II objects, according to 
the criteria of \citet{2012ApJ...744..130K}.} The other three (IRS 4, 5 and 6)
have no mid-infrared counterparts and are likely younger objects \citep{2009ApJS..181..360C}. We model
the SEDs in the range 20--1200 $\mu$m of the four brightest sources in these wavelengths (IRS 1, 4, 5 and 6), 
deriving physical parameters for the associated envelopes,
such as the envelope masses, sizes, densities, and luminosities.  These parameters range from 16
to 68 $M_{\sun}$, 0.06\,--\,0.12 pc, 9.6$\times$10$^{4}$\,--\, 9.2$\times$10$^{6}$ cm$^{-3}$,
and 0.12\,--\,2.6$\times$10$^{3}$\,\,$L_{\sun}$, suggesting that these sources are very young,
massive and luminous objects at early stages of the formation process. 

We constructed the color-color diagrams in the bands of WISE and 2MASS to identify potential  
young objects in the region. Those identified in the bands of WISE
are contaminated by the emission of PAHs, except four sources (IRS\,\,1, 2, 3 and 7). 
These faint candidates require more sensitive observations to confirm
their nature and evolutionary status. 

In the WISE 12 $\mu$m band, we identify an arc-like structure linked to {IRS}\,\,1, pointing to the west,
and other two smaller structures coinciding with the positions of other two infrared sources (IRS 5 and 6)
lying in the region, with similar shapes and pointing in the same direction. 
The massive star-forming region RWC 38,
harboring a dozen of O-type stars, is located $\sim$ 16\farcm 7 from these sources in the west
direction and roughly at the same distance. The ultraviolet photon flux from the exciting stars of
RCW 38 is probably photodissociating the material in the IRAS 08589$-$4714 region and creating a
photodissociated region.

\begin{acknowledgements}
H. P. S. acknowledges financial support from a fellowship from CONICET.
This project was partially financed by CONICET of Argentina under projects PIP 00356, and PIP 00107
and from UNLP, projects PPID092, PPID/G002, and 11/G120. M.R. wishes to acknowledge
support from CONICYT (CHILE) through FONDECYT grant No 1140839.
\end{acknowledgements}

\end{document}